\documentclass[conference, 10pt, onecolumn]{IEEEtran}
\usepackage{amsmath}
\usepackage[dvips]{graphicx}
\usepackage{amsmath}
\usepackage{amssymb}
\usepackage{epsfig}
\usepackage{graphicx}
\usepackage{dsfont}
\usepackage{float}
\usepackage[lined,boxed,commentsnumbered]{algorithm2e}
\usepackage{xfrac}
\usepackage{color}
\usepackage{psfrag}
\usepackage{amsthm}
\usepackage{subfigure}
\usepackage[bottom]{footmisc}
\usepackage{enumerate}
\usepackage{flushend}
\usepackage{caption}

\SetAlCapSkip{1em}

\DeclareMathAlphabet{\bm}{OML}{cmr}{bx}{it}

\newcommand{\mr}[1]{#1}
\newcommand{\bs}[1]{\bm{#1}}

\newcommand{\bg}[1]{\boldsymbol #1} %for bold greek variables

\newcommand{\bb}{\mathbb}
\newcommand{\rs}{\mathrm}
\newcommand{\mc}{\mathcal}

\newtheorem{lemma}{Lemma}
\newtheorem{proposition}{Proposition}

\newtheorem{remark}{Remark}

\newif\ifcomments
\commentsfalse

\usepackage[textsize=tiny]{todonotes}
\usepackage[color]{changebar}
\usepackage[normalem]{ulem}
\cbcolor{red}

\ifcomments
\newcommand{\removed}[1]{\cbstart\removedfragile{#1}\cbend{}}
\newcommand{\removedfragile}[1]{{\color{red}{\sout{#1}}}{}}
\newcommand{\added}[1]{\cbstart\addedfragile{#1}\cbend{}}

\newcommand{\changed}[2]{\added{#1}\removed{#2}}

\newcommand{\cosl}[1]{\todo[author=Slawomir,inline,color=yellow!40,caption={}]{#1}}
\else
\newcommand{\added}[1]{#1}
\newcommand{\removed}[1]{}
\newcommand{\changed}[2]{\added{#1}}
\newcommand{\cosl}[1]{}
\fi

\begin{document}
%
% paper title
% can use linebreaks \\ within to get better formatting as desired
\title{On $\ell_p$-norm Computation over Multiple-Access Channels}

% author names and affiliations
% use a multiple column layout for up to three different
% affiliations
\author{
\IEEEauthorblockN{Steffen Limmer\IEEEauthorrefmark{1},
		S\l awomir
                Sta\'nczak\IEEEauthorrefmark{1}\IEEEauthorrefmark{2}
                %Mario Goldenbaum\IEEEauthorrefmark{2},
		%Renato L. G. Cavalcante \IEEEauthorrefmark{1}
		}
		\IEEEauthorblockA{
	\IEEEauthorrefmark{1}
		Fraunhofer Institute for Telecommunications, Heinrich Hertz Institute,
		Einsteinufer 37,  10587 Berlin, Germany,\\
    \IEEEauthorrefmark{2}
    Fachgebiet Informationstheorie und theoretische Informationstechnik,\\
    Technische Universit\"at Berlin, Einsteinufer 25, 10587 Berlin, Germany\\		
	 %Email: \{steffen.limmer, slawomir.stanczak, renato.cavalcante\}@hhi.fraunhofer.de; mario.goldenbaum@tu-berlin.de
    }
}

% conference papers do not typically use \thanks and this command
% is locked out in conference mode. If really needed, such as for
% the acknowledgment of grants, issue a \IEEEoverridecommandlockouts
% after \documentclass

% use for special paper notices
\IEEEspecialpapernotice{(Invited Paper)}

% make the title area
\maketitle

\begin{abstract}
%\boldmath
This paper addresses some aspects of the general problem of
information transfer and distributed function computation in
wireless networks. Many applications of wireless technology foresee
networks of autonomous devices executing tasks that can be posed as
distributed function computation. In today's wireless networks, the
tasks of communication and (distributed) computation are performed
separately, although an efficient network operation calls for
approaches in which the information transfer is dynamically adapted
to time-varying computation objectives. Thus, wireless
communications and function computation must be tightly coupled and
it is shown in this paper that information theory may play a crucial
role in the design of efficient computation-aware wireless
communication and networking strategies. This is explained in more
detail by considering the problem of computing $\ell_p$-norms over
multiple access channels.
\end{abstract}

\IEEEpeerreviewmaketitle

%***************************************************************************
%***************************************************************************
%***************************************************************************
\section{Introduction}
\label{sec:intro}

Future wireless networks are envisioned \changed{to consist of}{as} a
massive \changed{large}{amount} of communication devices that perform
network tasks autonomously. A main enabler for this vision is the
ability of the network to \added{extract the useful information from}
\changed{a huge amount}{cope with large amounts} of data
\changed{distributed over the different nodes.}{containing only a
  considerably smaller amount of valuable information which is shared
  among the nodes.} In many scenarios, the network reveals its true
purpose-centric character, and the individual transmission of every
collected sensor value to some sink node can be circumvented. Consider
for example an environment monitoring system, where the main
\changed{objective}{purpose} is to make predictions \changed{with
  respect to}{about} a small set of state variables that are formed by
aggregating measured values of the network nodes. In such a setting,
the wireless network can leverage a channel that emerges from a dumb
bit-pipe to a network and signal processing building block providing
arithmetic operations that are facilitated by the laws of physics. In
particular, the superposition property of electromagnetic waves
provides certain computation capabilities inherently. In turn, this
property can drive the convergence between pure transmission of
waveforms and performing arithmetic operations on network
variables. In this regard, \changed{References}{the authors}
\cite{NaGa07,GoSt13} showed that channel collisions can be exploited
through a generalized \emph{Computation over Multiple-Access Channels
  (CoMAC)} framework. This framework subsumes techniques and
mechanisms for coding and transmission to compute functions at a
designated sink node using the superposition property of the
multiple-access channel. In \cite{GoSt13,goldenbaum2010computing}, the
authors showed that natural characteristics allow to compute functions
contained in the space of \emph{nomographic functions}. Continuing the
analysis of computable functions within this framework, this work
analyzes the computation of $\ell_p$ norms. Computing $\ell_p$ norms
is of high practical relevance for many applications, as it allows to
compute the number of non-zero elements for $p \to 0$ (and various
proxies for $0 < p \leq 1$) or the maximum value for $p \to \infty$
\added{(see Fig. \ref{fig:lp_norms})}. The main contribution of this
work is a unified analysis for \emph{CoMAC} of $\ell_p$ using short
sequences and a fixed energy detector at the
receiver.

%***************************************************************************
%***************************************************************************
\subsection{Notation}

Scalars, vectors and matrices are denoted by lowercase, bold lowercase
and bold uppercase letters, respectively. $\bb{R}$, $\bb{R}_{+}$, $\bb{Z}_+$ and $\bb{N}$ denote the
sets of real, nonnegative real, nonnegative integer and
natural numbers. $\bold{0}$, $\bold{1}$ and $\bm{I}_K$ denote the
vector of all zeros, all ones and the identity matrix of size $K
\times K$. $\rs{tr}\{\cdot\}$, $\rs{vec}\{\cdot\}$ and
$\rs{unvec}\{\cdot\}$ denote the trace of a matrix, the vectorization
of a matrix obtained by stacking it's columns and the inverse
vectorization operation. $\mc{N}(\cdot,\cdot)$ and $\bb{E}\left[ \cdot
\right]$ describe the normal distribution and the expectation
operator. $(\cdot)^T$ and $\otimes$ denote transposition and kronecker
product, respectively.

%********************************************************************************
%********************************************************************************
%********************************************************************************
\section{Computing $\ell_p$-norms over Multiple-Access Channels}
\label{sec:model}

\begin{figure}
\centering
% generated by laprint.m
%
\begin{psfrags}%
\psfragscanon%
%
% text strings:
\psfrag{s05}[t][t]{\color[rgb]{0,0,0}\setlength{\tabcolsep}{0pt}\begin{tabular}{c} {$x$} \end{tabular}}%
\psfrag{s06}[b][b]{\color[rgb]{0,0,0}\setlength{\tabcolsep}{0pt}\begin{tabular}{c} {$\lvert x \rvert^p$} \end{tabular}}%
\psfrag{s10}[][]{\color[rgb]{0,0,0}\setlength{\tabcolsep}{0pt}\begin{tabular}{c} \end{tabular}}%
\psfrag{s11}[][]{\color[rgb]{0,0,0}\setlength{\tabcolsep}{0pt}\begin{tabular}{c} \end{tabular}}%
\psfrag{s12}[l][l]{\color[rgb]{0,0,0}\footnotesize{p=2}}%
\psfrag{s13}[l][l]{\color[rgb]{0,0,0}\footnotesize{p=.25}}%
\psfrag{s14}[l][l]{\color[rgb]{0,0,0}\footnotesize{p=.5}}%
\psfrag{s15}[l][l]{\color[rgb]{0,0,0}\footnotesize{p=1}}%
\psfrag{s16}[l][l]{\color[rgb]{0,0,0}\footnotesize{p=2}}%
%
% xticklabels:
\psfrag{x01}[t][t]{\footnotesize{-1.5}}%
\psfrag{x02}[t][t]{\footnotesize{-1}}%
\psfrag{x03}[t][t]{\footnotesize{-0.5}}%
\psfrag{x04}[t][t]{\footnotesize{0}}%
\psfrag{x05}[t][t]{\footnotesize{0.5}}%
\psfrag{x06}[t][t]{\footnotesize{1}}%
\psfrag{x07}[t][t]{\footnotesize{1.5}}%
%
% yticklabels:
\psfrag{v01}[r][r]{\footnotesize{0}}%
\psfrag{v02}[r][r]{\footnotesize{0.5}}%
\psfrag{v03}[r][r]{\footnotesize{1}}%
\psfrag{v04}[r][r]{\footnotesize{1.5}}%
\psfrag{v05}[r][r]{\footnotesize{2}}%
\psfrag{v06}[r][r]{\footnotesize{2.5}}%
%
% Figure:
%\parbox{8cm}{\centering%
%\resizebox{8cm}{!}
\centering

\includegraphics[width= 0.5 \linewidth]{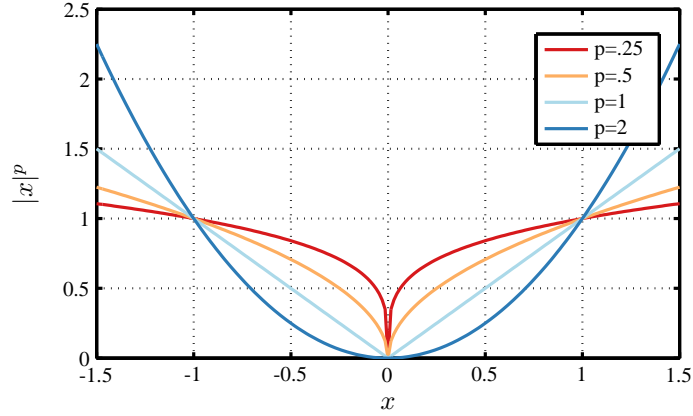}%
\caption{$\ell_p$ (pseudo-)norms for different values of $p$.}%
\label{fig:lp_norms}%

\end{psfrags}%
%
% End lp_norms.tex

\end{figure}

\begin{figure}[htb]
\centering
 \psfrag{x1}[bc][bc]{$\bm{x}_1$}
 \psfrag{xK}[bc][bc]{$\bm{x}_K$}
 \psfrag{p}[Bc][Bc]{${\varphi}(\cdot)$}
 \psfrag{s1}[Bc][Bc]{$\bm{s}_1$}
 \psfrag{sK}[Bc][Bc]{$\bm{s}_K$}
 \psfrag{n}[bc][bc]{$\bm{n}$}
 \psfrag{N}[bc][bc]{$\lVert \cdot \rVert_2^2$}
 \psfrag{fh}{$\hat{f}(\bm{x})$}
  \includegraphics[width= 0.5\linewidth]{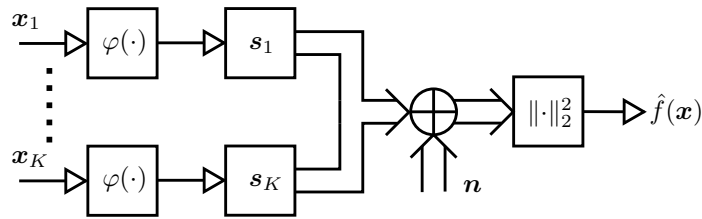}
 \caption{System structure of the proposed $\ell_p$-norm computation scheme.}
\label{fig:sys1}
\end{figure}

Consider a wireless sensor network consisting of \added{one designated
  sink node and} $K \in \bb{N}$ nodes that monitor a physical quantity
by sensor values $[x_1,\hdots,x_K]^T:=\bm{x} \in \bb{R}^K$\removed{and
  one designated sink node}. The objective of the network is to
compute functions of the form
\begin{align}\label{equ:ellp}
f(\bm{x}) = \lVert \bm{x} \rVert_p^p = \sum_{k=1}^K \lvert x_k \rvert^p
\end{align}
at the\removed{designated} sink node \changed{for given values of
  $p>0$ (see also Fig. \ref{fig:lp_norms} for illustration)}{, which
  are depicted in Fig. \ref{fig:lp_norms} for certain values of $p$.}
Precisely, the class of functions to be computed (referred to as
desired functions in what follows \added{in accordance with
  \cite{GoSt13}}) is given by the $\ell_p$-(pseudo)-norm to the power
of $p$.\footnote{Here, the prefix \emph{pseudo} applies to $p < 1$, in
  which case the functions are not a proper norm.}  In order to
compute such non-linear functions, we assume \added{as in
  \cite{GoSt13}} that \changed{each node is}{nodes are} equipped with
\added{a} pre-processing unit\removed{s} ${\varphi}:\bb{R}
\to \bb{R}_{+}$ and transmit their \emph{pre-processed} values
simultaneously to the sink node using transmit sequences
$\bm{S}=[\bm{s}_1, \hdots, \bm{s}_K] \in \bb{R}^{M \times K}$ (see
Fig. \ref{fig:sys1}).  For ease of exposition, we\removed{will} assume
in the following \changed{that}{(1) that}
\begin{enumerate}[(i)]
\item the sensor nodes have perfect channel knowledge so that the effect of
the communication channel can be perfectly equalized,
\item \removed{(2)} the nodes can be synchronized on frame and symbol
level and 
\item \removed{(3)} the receiver side detector is restricted to pure
  energy detection.\footnote{For an analysis of different channel
    state information schemes at the sensor nodes and the effect of
    coarse frame synchronization we refer the reader to
    \cite{GoSt13}. An analysis of more complex (e.g. affine or
    nonlinear) receiver-side detectors is beyond the scope of the
    paper.}
\end{enumerate}
Moreover, we treat the problem in the real-valued domain and
\changed{point out that}{note, that} the analysis for the complex
domain follows along similar lines.
\added{
\begin{remark}
  \label{rem:unconstrained_seq}
  It is important to emphasize that except for the requirement of
  real-valued entries, there are no additional constraints on the
  matrix $\bm{S}$. This means that transmit sequences are jointly
  optimized with transmit powers.  This stands in contrast to the
  studies related to CDMA systems, where the spreading sequences are
  normalized to be of unit norm and transmit powers are defined
  separately (see for instance \cite{boche2002iterative}).
\end{remark}
}

In the proposed setup depicted in Fig. \ref{fig:sys1}, the received
signal $\bm{y}$ and the output of the energy detector $\hat{f}$
are\added{, respectively,} given by
\begin{align}
\bm{y} & = \bm{S} \bg{\varphi}(\bm{x}) + \bm{n} \\
\label{equ:defs1}
\hat{f} & = \lVert \bm{S} \bg{\varphi}(\bm{x}) + \bm{n} \rVert_2^2\,.
\end{align}
\removed{respectively.}Here, $\bg{\varphi}: \bb{R}^K \to \bb{R}^K$
denotes the concatenated mapping from raw sensor readings $\bm{x}$ to
transmit symbols $\bg{\varphi}(\bm{x})$ that is defined to be
\begin{align}
\bg{\varphi}(\bm{x}) := [\lvert x_1 \rvert^{\frac{p}{2}}, \hdots, \lvert x_K \rvert^{\frac{p}{2}}]^T.
\end{align}
\begin{remark}\label{rem:ideal1}
In an idealized setting with $\bm{n}=\bold{0}$ and $M \geq K$, the desired function value can be recovered exactly using any $\bm{S}$ such that $\bm{S}^T \bm{S} = \bm{I}_K$. In this case, we have
\begin{align}\label{equ:idf}
\hat{f}(\bm{x}) & = \lVert \bm{S} \bg{\varphi}(\bm{x}) \rVert_2^2 = \bg{\varphi}(\bm{x})^T \bm{S}^T \bm{S} \bg{\varphi}(\bm{x}) = \sum_{k=1}^K \left( \vert x_k \rvert^{\frac{p}{2}} \right)^2 \\
& = \lVert \bm{x} \rVert_p^p \equiv f(\bm{x}) \nonumber.
\end{align}
\end{remark}
However, the case described in Remark \ref{rem:ideal1} is of minor
practical relevance, as it excludes the effect of noise and is limited
to the case $K\geq M$, i.e. sequence lengths that exceed the number of
network nodes.  The more interesting case is a noisy setting with
$M<K$, and we motivate the use of the given pre-processing function
and system structure by the identity in \eqref{equ:idf}. We note that
in addition to the application for networked computation of
$\ell_p$-norms, the considered setting is also closely related to
problems in coding theory and robust dimensionality reduction (see
e.g. \cite{StHe03},\cite{LiHaCh07}).  

To simplify the subsequent analysis, we fix some $M < K$ and consider
an \emph{MSE} metric for estimating $f$ by $\hat{f}$ using $\bm{S}$
and a fixed receiver structure as depicted in Fig. \ref{fig:sys1}. In
this case, the MSE can be computed as
\begin{align}\label{equ:J0}
& J(\bm{S}) = \bb{E}\left[ (f - \hat{f})^2 \right] \\
& = \bb{E} \left[ \left( \underbrace{\bg{\varphi}(\bs{x})^T ( \bm{I}_K - \bm{S}^T\bm{S} )  \bg{\varphi}(\bs{x})}_{a} - \underbrace{2 \bg{\varphi}(\bs{x})^T \bm{S}^T \bs{n}}_{b} - \underbrace{\bs{n}^T \bs{n}}_{c} \right)^2 \right] \nonumber \\
& = \bb{E} \left[ a^2 + b^2 + c^2 -2ab -2ac + 2bc \right], \nonumber
\end{align}
which can be evaluated assuming that the probability distribution
functions of $\bs{x}$ and $\bs{n}$ are given.
% \footnote{The proposed framework can also be applied directly for
% other isotropic distributions provided the involved expectations
% exist. An interesting example is a uniform spherical prior on
% $\bs{x}$, where the required moments are given in \cite{Ba97}.}
\begin{lemma}\label{lem:J0}
  Assume that $\bs{x}$ and $\bs{n}$ are independent and distributed as
  $\bs{x}~\sim~\mc{N}(\bold{0},\sigma_{\mr{x}}^2 \bm{I}_K)$ and
  $\bs{n}~\sim~\mc{N}(\bold{0}, \sigma_{\mr{n}}^2 \bm{I}_M)$.  Under
  this assumption, the expectation in \eqref{equ:J0} decomposes and
  \changed{we have}{the terms yield}
\begin{align}\label{equ:exp1}
\bb{E} \left[ a^2 \right] & = \rs{tr}\left\{   \bm{M} (\bm{S}^T \bm{S} \otimes \bm{S}^T \bm{S} )  \right\}  \\
& - 2 \rs{tr} \left\{ \bm{M} ( \bm{I} \otimes \bm{S}^T \bm{S} )  \right\} + \rs{tr} \left\{ \bm{M} \right\} \nonumber \\
\bb{E} \left[ b^2 \right] & = 4 \sigma_{\mr{n}}^2 \rs{tr} \left\{ \bm{C} \bm{S}^T \bm{S} \right\} \nonumber \\
\bb{E} \left[ c^2 \right] & = \rs{tr} \{ \bm{N} \} \nonumber \\
\bb{E} \left[ a c \right] & = M \sigma_{\mr{n}}^2 \rs{tr} \left\{ \bm{C} (\bm{I} - \bm{S}^T \bm{S}) \right\} \nonumber \\
\bb{E}\left[ab \right] & = \bb{E} \left[ bc \right] = 0 \nonumber,
\end{align}
with
\begin{align}\label{equ:exp2}
\bm{C} & = \bb{E} \left[ \bg{\varphi}(\bs{x}) \bg{\varphi}(\bs{x})^T \right] \\
\bm{M} & = \bb{E} \left[ \rs{vec} \left\{ \bg{\varphi}(\bs{x}) \bg{\varphi}(\bs{x})^T \right\} \rs{vec} \left\{ \bg{\varphi}(\bs{x}) \bg{\varphi}(\bs{x})^T \right\}^T \right] \nonumber \\
\bm{N} & = \bb{E} \left[ \rs{vec}\left\{ \bs{n} \bs{n}^T\right\} \rs{vec}\left\{ \bs{n} \bs{n}^T \right\}^T \right] \nonumber,
\end{align}
\end{lemma}
\begin{proof}
The result for $\bb{E}[a^2]$ follows from 
\begin{align}
& \bg{\varphi}(\bs{x})^T \bm{A} \bg{\varphi}(\bs{x}) \bg{\varphi}(\bs{x})^T \bm{B} \bg{\varphi}(\bs{x}) = \\
& = \rs{tr} \left\{ \bg{\varphi}(\bs{x}) \bg{\varphi}(\bs{x})^T \bm{A} \bg{\varphi}(\bs{x}) \bg{\varphi}(\bs{x})^T \bm{B} \right\} \nonumber \\
& = \rs{vec} \left\{\bm{A}^T \bg{\varphi}(\bs{x})\bg{\varphi}(\bs{x})^T \right\}^T \rs{vec} \left\{\bg{\varphi}(\bs{x})\bg{\varphi}(\bs{x})^T \bm{B} \right\} \nonumber \\
% & = \rs{tr} \left\{ \left[ ( \bold{1} \otimes \bm{A}^T) \rs{vec} \left\{ \bg{\varphi}(\bs{x})\bg{\varphi}(\bs{x})^T \right\} \right]^T  (\bm{B}^T \otimes \bold{1}) \rs{vec} \left\{\bg{\varphi}(\bs{x})\bg{\varphi}(\bs{x})^T \right\} \right\} \nonumber \\
% &= \rs{tr} \left\{ (\bold{1} \otimes \bm{A}) (\bm{B}^T \otimes \bold{1}) \rs{vec} \left\{\bg{\varphi}(\bs{x})\bg{\varphi}(\bs{x})^T\right\} \rs{vec} \left\{\bg{\varphi}(\bs{x})\bg{\varphi}(\bs{x})^T\right\}^T \right\} \nonumber \\
& = \rs{tr} \left\{ (\bm{B}^T \otimes \bm{A}) \rs{vec} \left\{\bg{\varphi}(\bs{x})\bg{\varphi}(\bs{x})^T\right\} \rs{vec} \left\{\bg{\varphi}(\bs{x})\bg{\varphi}(\bs{x})^T\right\}^T \right\} \nonumber
\end{align}
using 
\begin{align}
\rs{tr}\left\{\bm{A} \bm{B} \bm{C} \right\} & = \rs{tr}\left\{\bm{C}  \bm{A} \bm{B} \right\} = \rs{tr}\left\{\bm{B} \bm{C}  \bm{A}  \right\} \\
\rs{tr}\left\{\bm{A}^T \bm{B} \right\} & = \rs{vec}\left\{\bm{A}\right\}^T \rs{vec}\left\{\bm{B} \right\} \nonumber \\
\rs{vec}\left\{ \bm{A} \bm{B} \bm{C} \right\} & = ( \bm{C}^T \otimes \bm{A} ) \rs{vec}\left\{\bm{B}\right\} \nonumber \\
(\bm{A} \otimes \bm{B})(\bm{C} \otimes \bm{D}) &= \bm{A}\bm{C} \otimes \bm{B} \bm{D}, \nonumber
\end{align}
and performing the expectation w.r.t. the random variable
$\bs{x}$. The term $\bb{E}[c^2]$ follows similarly with $\bs{n}$ as
random variable. $\bb{E}[b^2]$, $\bb{E}[ac]$, $\bb{E}[ab]$ and
$\bb{E}[b^2]$ follow from independency of $\bs{x}$ and $\bs{n}$ and
the zero mean assumption on $\bs{n}$, respectively.
\end{proof}

\added{Moreover, we point out the following without a proof. 
\begin{lemma}
\label{lem:min_exists}
The MSE function $J:\bb{R}^{M\times K}\to\bb{R}_{+}$ defined by
(\ref{equ:J0}) attains a minimum on $\mathbb{R}^{M\times K}$.
\end{lemma}
}

To compute the matrices $\bm{C}$, $\bm{M}$ and $\bm{N}$, we note, that
all matrix entries are equal to monomials with either non-negative
rational exponents $\bg{\alpha} \in \bb{R}_{+}^K$ for $\bm{C}$ and
$\bm{M}$, or non-negative integer exponents $\bg{\beta} \in
\bb{Z}_+^M$ for $\bm{N}$.  Thus, the entries of $\bm{C}$ and $\bm{M}$
can be computed by extending the derivation of \emph{central absolute
  moments} in \cite{Wi12} to the monomial case:
\begin{align}
Q(\bg{\alpha}) & = \bb{E}\left[ \prod_{k=1}^K \lvert {{\mr{x}}_k} \rvert ^{\alpha_k}  \right] 
 = \int_{\bb{R}^K} \lvert {x}_1 \rvert ^{\alpha_1} \cdots \lvert {x}_K \rvert ^{\alpha_K} p_\bs{x}(\bm{x}) d \bm{x}  \nonumber \\
 & = \frac{(2 \sigma_{\mr{x}}^2)^{\sum_{k=1}^K \frac{\alpha_k}{2}}}{\sqrt{\pi}^K} \prod_{k=1}^K \Gamma \left( \frac{\alpha_k +1}{2} \right).
\end{align}
Similarly, the entries of $\bm{N}$ can be obtained by extending the
derivation of \emph{central moments} in \cite{Wi12}:
\begin{align}
 I(\bg{\beta}) & = \bb{E}\left[ \prod_{m=1}^M {\mr{n}}_k^{\beta_k} \right] 
 = \int_{\bb{R}^M} n_1^{\beta_1} \cdots n_M^{\beta_M} p_\bs{n}(\bm{n}) d \bm{n} \\
& = \begin{cases}
0 & \text{if some $\beta_m$ is odd} \\
\frac{(2 \sigma_{\mr{n}}^2)^{\sum_{m=1}^M \frac{\beta_m}{2}}}{\sqrt{\pi}^M} \prod_{m=1}^M \Gamma( \frac{\beta_m +1}{2}) &  \text{if all $\beta_m$ are even.} \nonumber
\end{cases}
\end{align}

\added{Now we are in a position to state our optimization problem.}

\begin{proposition}
  \label{prop:opt1}
Let $\hat{f}$ be given by \eqref{equ:defs1},
  \added{and let} $\bs{x}~\sim~\mc{N}(\bold{0},\sigma_{\mr{x}}^2
  \bm{I}_K)$ and $\bs{n}~\sim~\mc{N}(\bold{0},\sigma_{\mr{n}}^2
  \bm{I}_M)$ be independent. Then, the optimal sequences for computing
  $f=\lVert x \rVert_p^p$ given the receiver structure in
  Fig. \ref{fig:sys1} w.r.t. an MSE metric can be obtained as
  \changed{a}{the} solution to the problem
\begin{align}
  \label{equ:opt1}
  \underset{\substack{\bm{S} \in \bb{R}^{M \times K}}}{\operatorname{min}} \ & \rs{tr}\left\{ \bm{M} (\bm{S}^T \bm{S} \otimes \bm{S}^T \bm{S}) \right\} - 2\rs{tr} \left\{ \bm{M} (\bold{1} \otimes \bm{S}^T\bm{S}) \right\} \nonumber \\
  & + \rs{tr}\left\{\bm{M} \right\} + 4 \sigma_{\mr{n}}^2 \rs{tr}\left\{ \bm{C} \bm{S}^T \bm{S} \right\} \nonumber \\
  & + \rs{tr} \left\{\bm{N} \right\} -2M \sigma_{\mr{n}}^2 \rs{tr}\left\{ \bm{C}\left( \bm{I} - \bm{S}^T\bm{S} \right) \right\} \\
  = \underset{\substack{\bm{S} \in \bb{R}^{M \times K}}}{\operatorname{min}} \
  & J(\bm{S})\,.\nonumber
\end{align}
\end{proposition}
\begin{proof}
  The proof follows directly by combining \added{Lemma
    \ref{lem:min_exists}}, Lemma \ref{lem:J0} and \eqref{equ:J0}.
\end{proof}
To the best of our knowledge, a closed-form solution to this problem
is not known. Consequently we \changed{are going to}{will} approach
the problem by numerical methods.

%********************************************************************************
%********************************************************************************
%********************************************************************************
\section{First-Order Optimization of Transmit Sequences}
\label{sec:opt_sequences}

In this section, we develop a simple gradient descent algorithm to
optimize the cost function $J$ over the unconstrained input domain
$\bb{R}^{M \times K}$. \added{Unfortunately, the problem of
  Proposition \ref{prop:opt1} (Problem (\ref{equ:opt1})) is not convex
  and there is no guarantee that the algorithm converges to a global
  minimum. The problem of designing an algorithm with global
  convergence is left as an open problem for future research.}

\changed{Instead}{To this end}, we \changed{make use of}{refer to} the
well-known gradient descent iteration \added{to optimize the MSE
  function $J$:}
\begin{align}
\bm{S}^{(t+1)} = \bm{S}^{(t)} - \mu^{(t)} \nabla_\bm{S} J(\bm{S}^{(t)}),
\end{align}
where a suitable step-size that guarantees a non-increasing sequence
of objective values $J(\bm{S}^{(t)})$ can be obtained using the
Armijo criterion \cite{WrNo99,MiMeSe11}
\begin{align}\label{equ:armijo}
J \left( \bm{S}^{(t)} - \mu^{(t)} \nabla_\bm{S} J(\bm{S}^{(t)}) \right) \leq J(\bm{S}^{(t)}) - c \mu^{(t)} \lVert \nabla_\bm{S} J(\bm{S}^{(t)}) \rVert_F^2.
\end{align}
\added{In fact, it can be shown that a \emph{sufficiently small}
  constant step size would guarantee a non-increasing sequence of
  objective values as well. Since the objective function is bounded
  below (it is greater than zero), we can conclude that the sequence
  $J(\bm{S}^{(t)})$ must converge under (\ref{equ:armijo}). This fact
  will be used for termination condition.}

To obtain an analytic expression for $\nabla_\bm{S}J(\bm{S})$ we refer
to the analytic expression in Lemma \ref{lem:J0} and \eqref{equ:J0},
and simple standard formulas (e.g. \cite{HjGe07}) yield
\begin{align}\label{equ:grads}
\nabla_\bm{S} \bb{E}\left[ b^2 \right] & = 4 \sigma_{\mr{n}}^2 ( \bm{S} \bm{C}^T + \bm{S} \bm{C} ) \\
\nabla_\bm{S} \bb{E}\left[ ac \right] & = - M \sigma_{\mr{n}}^2 ( \bm{S} \bm{C}^T + \bm{S} \bm{C} ) \nonumber \\
\nabla_\bm{S} \bb{E}\left[ b^2 \right] & = \nabla_\bm{S} \bb{E}\left[ c^2 \right] = 0. \nonumber
\end{align}
It remains to compute $\nabla_\bm{S} \bb{E}\left[ a^2 \right]$. To this end, we introduce the following lemma:
\begin{lemma}\label{lem:ksvd}
Let $\bm{A} \in \bb{R}^{K^2 \times K^2}$ and $\mc{R}(\bm{A})$ denote a permutation of $\bm{A}$ given by 
\begin{align}
\mc{R}(\bm{A}) & = \left[ \rs{vec}\left\{ \bm{A}_{1,1} \right\}, \hdots, \rs{vec}\left\{ \bm{A}_{K,1} \right\}, \rs{vec}\left\{ \bm{A}_{1,2} \right\}, \hdots, \right. \nonumber \\
& \left. \rs{vec}\left\{ \bm{A}_{1,2} \right\}, \hdots, \rs{vec}\left\{ \bm{A}_{K,K} \right\} \right]^T
\end{align}
where $\bm{A}_{i,j} \in \bb{R}^{K \times K}$, $i,j\in \{1,\hdots,K\}$ denote the $K \times K$ block matrix partitioning of $\bm{A}$.
If $\mc{R}(\bm{A})$ has a singular value decomposition
\begin{align}
\mc{R}(\bm{A}) = \sum_{k=1}^{K^2} \sigma_k \bm{u}_k \bm{v}_k^T,
\end{align}
where $\sigma_k$, $\bm{u}_k$ and $\bm{v}_k$ are the corresponding singular values and singular vectors, then the matrices $\bm{U}_k = \rs{unvec}\{ \bm{u}_k \}$ and $\bm{V}_k= \rs{unvec} \{ \bm{v}_k \}$ form a decomposition
\begin{align}
\bm{A} = \sum_{k=1}^{K^2} \sigma_k \bm{U}_k \otimes \bm{V}_k.
\end{align}
\end{lemma}
\begin{proof}
The proof follows directly from Corollary 2.2 in \cite{VaPi93}.
\end{proof}
\begin{remark}\label{rem:kevd}
In our case $\bm{M}$ has the additional property that $\mc{R}(\bm{M})=\bm{M}=\bm{M}^T$, which is stated here without proof. Hence, the SVD in Lemma \ref{lem:ksvd} can be replaced by an EVD. Consequently, the matrix $\bm{M}$ can be decomposed as
\begin{align}\label{equ:krondec}
\bm{M} = \sum_{k=1}^{K^2} \bm{M}_k \otimes \bm{M}_k,
\end{align}
with $\bm{M}_k := \sqrt{\sigma_k} \bm{U}_k \equiv \sqrt{\sigma_k} \bm{V}_k$. 
\end{remark}
Now we are in a position to obtain an analytic expression for $\nabla \bb{E}[a^2]$.
\begin{proposition}\label{prop:dEa2}
Let $\bb{E}[a^2]$ be given according to Lemma \ref{lem:J0} and $\bm{M}$ according to Lemma \ref{lem:ksvd} and Remark \ref{rem:kevd}. Then, $\nabla \bb{E}[a^2]$ is given by
\begin{align}
\nabla_\bm{S} \bb{E}[a^2] &= 2 \sum_{k=1}^{K^2} \rs{tr} \{ \bm{M}_k \bm{S}^T \bm{S} \} \cdot ( \bm{S} \bm{M}_k + \bm{S} \bm{M}_k^T ) \\
& -2 \sum_{k=1}^{K^2} \rs{tr}\{\bm{M}_k\} \cdot ( \bm{S} \bm{M}_k^T + \bm{S} \bm{M}_k ). \nonumber
\end{align}
\end{proposition}
\begin{proof}
Define $\nabla_\bm{S} \bb{E}[a^2] := \Delta^{(1)} - 2\Delta^{(2)}+\Delta^{(3)}$ with
\begin{align}
\Delta^{(1)} & := \nabla_\bm{S} \rs{tr}\left\{ \bm{M} (\bm{S}^T \bm{S} \otimes \bm{S}^T \bm{S} ) \right\} \\
\Delta^{(2)} & := \nabla_\bm{S} \rs{tr} \left\{ \bm{M} ( \bm{I} \otimes \bm{S}^T \bm{S} ) \right\} \nonumber \\
\Delta^{(3)} & := \nabla_\bm{S}  \rs{tr} \left\{ \bm{M} \right\}. \nonumber
\end{align}
Using Lemma \ref{lem:ksvd}, Remark \ref{rem:kevd} and trace derivatives (e.g. \cite{HjGe07}) we obtain
\begin{align}\label{equ:Ea1}
\Delta^{(1)} & = \nabla_\bm{S} \sum_{k=1}^{K^2} \rs{tr}\left\{ (\bm{M}_k \otimes \bm{M}_k) \cdot (\bm{S}^T \bm{S} \otimes \bm{S}^T \bm{S} ) \right\}  \\
& = \nabla_\bm{S} \sum_{k=1}^{K^2}  \rs{tr}\{ \bm{M}_k \bm{S}^T \bm{S} \}  \rs{tr}\{ \bm{M}_k \bm{S}^T \bm{S} \} \nonumber \\
& = 2 \sum_{k=1}^{K^2} \rs{tr} \{ \bm{M}_k \bm{S}^T \bm{S} \} \cdot ( \bm{S} \bm{M}_k + \bm{S} \bm{M}_k^T ), \nonumber
\end{align}
\begin{align}\label{equ:Ea2}
\Delta^{(2)} & = \nabla_\bm{S} \sum_{k=1}^{K^2} \rs{tr}\left\{ (\bm{M}_k \otimes \bm{M}_k) \cdot (\bold{1} \otimes \bm{S}^T \bm{S} ) \right\} \\
& = \nabla_\bm{S} \sum_{k=1}^{K^2} \rs{tr}\{\bm{M}_k\} \cdot \rs{tr}\{ \bm{M}_k \bm{S}^T \bm{S} \} \nonumber \\
& = \sum_{k=1}^{K^2} \rs{tr}\{\bm{M}_k\} \cdot ( \bm{S} \bm{M}_k^T + \bm{S} \bm{M}_k ), \nonumber
\end{align}
and
\begin{align}
\Delta^{(3)} & = 0,
\end{align}
which completes the proof. 
\end{proof}
Using Proposition \ref{prop:dEa2} and \eqref{equ:grads}, the overall gradient of $J(\bm{S})$ w.r.t. $\bm{S}$ is given by
\begin{align}
\nabla_\bm{S} J(\bm{S}) &= \nabla_\bm{S} \bb{E}[a^2] + \nabla_\bm{S} \bb{E}[b^2] - 2 \nabla_\bm{S} \bb{E}[ac] \\
& = \Delta^{(1)} - 2\Delta^{(2)} + (2M+4) \sigma_{\mr{n}}^2  \left(\bm{S} \bm{C}^T + \bm{S} \bm{C}\right). \nonumber
\end{align}
In practice, the sum involved in the computation of $\Delta^{(1)}$ and $\Delta^{(2)}$ can be truncated depending on the decay of singular values $\sigma_k$ to reduce the computational burden.
The resulting gradient descent algorithm is described in Alg. \ref{alg:alg1}.
\begin{algorithm}[h]
\caption{Gradient descent algorithm for optimizing sequence matrices.}
\label{alg:alg1}
\KwData{initial iterate $\bm{S}^{(0)}$, desired norm $p$}
\KwResult{optimized matrix $\tilde{\bm{S}} \in \bb{R}^{M \times K}$}
initialization: compute $\bm{C}$, $\bm{M}$, $\bm{M}_k$, $\bm{N}$\;
\While{ $J(\bm{S}^{(t)}) - J(\bm{S}^{(t-1)}) \geq \varepsilon J(\bm{S}^{(t)}) $ {and} $t \leq T$}{
find $\mu^{(t)}$ that satisfies \eqref{equ:armijo}  \;
$\bm{S}^{(t+1)} = \bm{S}^{(t)} - \mu^{(t)} \nabla_{\bm{S}} J(\bm{S})$ \;
}
\end{algorithm}

\section{Numerical Results}
\begin{figure}
\centering
% This file is generated by the MATLAB m-file laprint.m. It can be included
% into LaTeX documents using the packages graphicx, color and psfrag.
% It is accompanied by a postscript file. A sample LaTeX file is:
%    \documentclass{article}\usepackage{graphicx,color,psfrag}
%    \begin{document}\input{sims_1e-3}\end{document}
% See http://www.mathworks.de/matlabcentral/fileexchange/loadFile.do?objectId=4638
% for recent versions of laprint.m.
%
% created by:           LaPrint version 3.16 (13.9.2004)
% created on:           08-Aug-2014 17:00:56
% eps bounding box:     15 cm x 9 cm
% comment:              
%
\begin{psfrags}%
\psfragscanon%
%
% text strings:
\psfrag{s05}[t][t]{\color[rgb]{0,0,0}\setlength{\tabcolsep}{0pt}\begin{tabular}{c}$p$\end{tabular}}%
\psfrag{s06}[b][b]{\color[rgb]{0,0,0}\setlength{\tabcolsep}{0pt}\begin{tabular}{c}$\log_{10}(J)$\end{tabular}}%
\psfrag{s10}[][]{\color[rgb]{0,0,0}\setlength{\tabcolsep}{0pt}\begin{tabular}{c} \end{tabular}}%
\psfrag{s11}[][]{\color[rgb]{0,0,0}\setlength{\tabcolsep}{0pt}\begin{tabular}{c} \end{tabular}}%
\psfrag{s12}[l][l]{\color[rgb]{0,0,0}\footnotesize{WBE: 6x16}}%
\psfrag{s13}[l][l]{\color[rgb]{0,0,0}\footnotesize {ALG 1: 3x6}}%
\psfrag{s14}[l][l]{\color[rgb]{0,0,0}\footnotesize {WBE: 3x6}}%
\psfrag{s15}[l][l]{\color[rgb]{0,0,0}\footnotesize {ALG 1: 6x16}}%
\psfrag{s16}[l][l]{\color[rgb]{0,0,0}\footnotesize {WBE: 6x16}}%
%
% xticklabels:
\psfrag{x01}[t][t]{0}%
\psfrag{x02}[t][t]{0.5}%
\psfrag{x03}[t][t]{1}%
\psfrag{x04}[t][t]{1.5}%
\psfrag{x05}[t][t]{2}%
\psfrag{x06}[t][t]{2.5}%
\psfrag{x07}[t][t]{3}%
\psfrag{x08}[t][t]{3.5}%
\psfrag{x09}[t][t]{4}%
%
% yticklabels:
\psfrag{v01}[r][r]{-5}%
\psfrag{v02}[r][r]{-4}%
\psfrag{v03}[r][r]{-3}%
\psfrag{v04}[r][r]{-2}%
\psfrag{v05}[r][r]{-1}%
\psfrag{v06}[r][r]{0}%
\psfrag{v07}[r][r]{1}%
\psfrag{v08}[r][r]{2}%
\psfrag{v09}[r][r]{3}%
%
% Figure:
\centering

\includegraphics[width= 0.5 \linewidth]{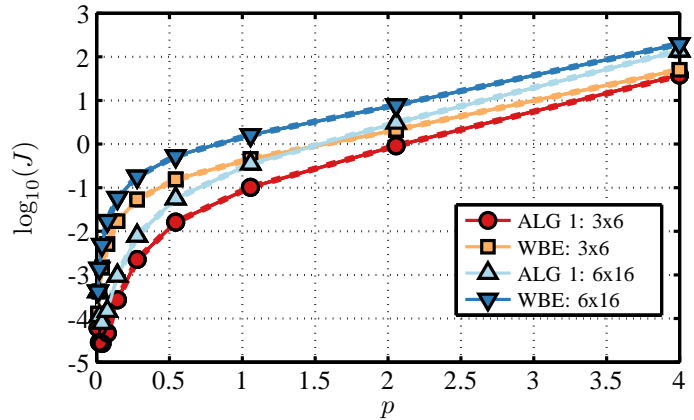}%
\caption{Simulation results for $\sigma_{\mr{x}}^2=1$ and $\sigma_{\mr{n}}^2=10^{-3}$. Monte Carlo results are shown in solid, analytical results from \eqref{equ:opt1} in dashed linestyle.}%
\label{fig:sims_1e-3}%
\end{psfrags}%
% End sims_1e-3.tex

\end{figure}

\begin{figure}
\centering
% This file is generated by the MATLAB m-file laprint.m. It can be included
% into LaTeX documents using the packages graphicx, color and psfrag.
% It is accompanied by a postscript file. A sample LaTeX file is:
%    \documentclass{article}\usepackage{graphicx,color,psfrag}
%    \begin{document}\input{sims_0.1}\end{document}
% See http://www.mathworks.de/matlabcentral/fileexchange/loadFile.do?objectId=4638
% for recent versions of laprint.m.
%
% created by:           LaPrint version 3.16 (13.9.2004)
% created on:           08-Aug-2014 13:33:41
% eps bounding box:     15 cm x 9 cm
% comment:              
%
\begin{psfrags}%
\psfragscanon%
%
% text strings:
\psfrag{s05}[t][t]{\color[rgb]{0,0,0}\setlength{\tabcolsep}{0pt}\begin{tabular}{c}$p$\end{tabular}}%
\psfrag{s06}[b][b]{\color[rgb]{0,0,0}\setlength{\tabcolsep}{0pt}\begin{tabular}{c}$\log_{10}(J)$\end{tabular}}%
\psfrag{s10}[][]{\color[rgb]{0,0,0}\setlength{\tabcolsep}{0pt}\begin{tabular}{c} \end{tabular}}%
\psfrag{s11}[][]{\color[rgb]{0,0,0}\setlength{\tabcolsep}{0pt}\begin{tabular}{c} \end{tabular}}%
\psfrag{s12}[l][l]{\color[rgb]{0,0,0}\footnotesize {WBE: 6x16}}%
\psfrag{s13}[l][l]{\color[rgb]{0,0,0}\footnotesize {ALG 1: 3x6}}%
\psfrag{s14}[l][l]{\color[rgb]{0,0,0}\footnotesize {WBE: 3x6}}%
\psfrag{s15}[l][l]{\color[rgb]{0,0,0}\footnotesize {ALG 1: 6x16}}%
\psfrag{s16}[l][l]{\color[rgb]{0,0,0}\footnotesize {WBE: 6x16}}%
%
% xticklabels:
\psfrag{x01}[t][t]{0}%
\psfrag{x02}[t][t]{0.5}%
\psfrag{x03}[t][t]{1}%
\psfrag{x04}[t][t]{1.5}%
\psfrag{x05}[t][t]{2}%
\psfrag{x06}[t][t]{2.5}%
\psfrag{x07}[t][t]{3}%
\psfrag{x08}[t][t]{3.5}%
\psfrag{x09}[t][t]{4}%
%
% yticklabels:
\psfrag{v01}[r][r]{-1}%
\psfrag{v02}[r][r]{-0.5}%
\psfrag{v03}[r][r]{0}%
\psfrag{v04}[r][r]{0.5}%
\psfrag{v05}[r][r]{1}%
\psfrag{v06}[r][r]{1.5}%
\psfrag{v07}[r][r]{2}%
\psfrag{v08}[r][r]{2.5}%
%
% Figure:
\centering

\includegraphics[width= 0.5 \linewidth]{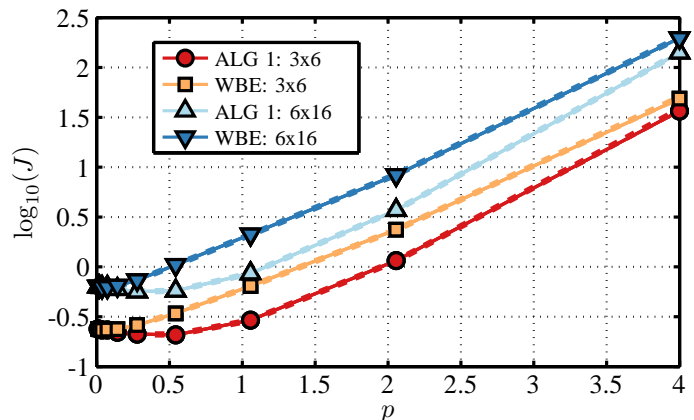}%
\caption{Simulation results for $\sigma_{\mr{x}}^2=1$ and $\sigma_{\mr{n}}^2=0.1$. Monte Carlo results are shown in solid, analytical results from \eqref{equ:opt1} in dashed linestyle.}%
\label{fig:sims_0.1}%
\end{psfrags}%
% End sims_0.1.tex

\end{figure}

To evaluate the performance of the proposed first-order optimization scheme we simulate a network consisting of $K \in \{6,16\}$ nodes and sequence lengths $M \in \{3,6\}$ to compute the desired function $f=\lVert \bm{x} \rVert_p^p$ for $p\in [10^{-3},4]$. The signal and noise powers are set to $\sigma_{\mr{x}}^2=1$ and $\sigma_{\mr{n}}^2 \in \{0.01,0.1\}$, the number of Monte Carlo iterations is $10^{5}$ and the gradient descent optimization is carried out using relative threshold $\varepsilon = 10^{-5}$, Armijo parameter $c=0.5$ and a maximum of $T=10^5$ gradient descent iterations. 
For comparison, we choose \emph{equiangular tight frames} (ETFs), which are known to meet both \emph{Welch Bound} and \emph{Maximum Welch Bound} with equality and are good candidate solutions for many applications in communications and coding (see e.g. \cite{StHe03}). For the simulations, we use scaled versions of the $3\times 6$ and $6 \times 16$ ETFs from \cite[p. 78f]{Re09}, where the scaling factor is obtained by line-search to optimize \eqref{equ:J0} (denoted by \emph{WBE}). The result is fed as initial iterate into the gradient descent optimization algorithm from Alg. \ref{alg:alg1} (denoted by \emph{ALG 1}). The results are depicted in Fig. \ref{fig:sims_1e-3} and \ref{fig:sims_0.1}. According to our simulation results, we can achieve considerable performance gains over ETFs for $p$ (strictly) between $10^{-3}$ and $4$. On the other hand, the performance gains for the case $p = 4$ and noise level $\sigma_{\mr{n}}^2=10^{-3}$ as well as the case $p = 10^{-3}$ and noise level $\sigma_{\mr{n}}^2=0.1$ are rather moderate.
\begin{remark}
It is important to emphasize that the transmit powers (norms) of the compared sequences \emph{are allowed to be different} in our setting. However, higher transmit powers do not necessarily result in a lower estimation error due to the fixed energy detector at the receiver (see Fig. \ref{fig:sys1}). In fact, our simulation results show, that optimized sequences can even have lower total/maximum transmit power in some cases.
\end{remark}


\section{Conclusion}
In this paper, we studied the problem of computing $\ell_p$~-~norms
over the wireless channel using a previously proposed scheme in an
idealized setting comprising perfect channel equalization and node
synchronization. Assuming a simple energy detection scheme at a
designated sink node and scalar pre-processing units at the
transmitter nodes we optimize sequences for the best MSE
performance. For the case of Gaussian priors on signal and noise, we
give a unified error-analysis for the resulting MSE as a function of
$p$ and the deployed transmit sequences. By using a simple gradient
descent scheme, we showed that \emph{(Maximum) Welch Bound Equality
  Sequences}, which are a good candidate solution for network tasks
involving interference avoidance, can be outperformed in terms of an
MSE criterion, though the performance gains in the investigated
small-scale network are rather moderate. An interesting direction to
further improve the MSE performance is the use of more complex
receiver structures as well as fixed-rank manifold based optimization
methods as outlined in \cite{MiMeSe11}. Promising applications of the outlined computation scheme involve
measuring the \emph{sparsity} of sensor values in a network or the
maximum sensor value. However, for measuring the sparsity of the
sensor values, the Gaussian signal prior poses a limitation in the
sense that typical realizations are not sparse. Using a more accurate
model for sparse signals by sparse processes or compressible
distributions constitutes an interesting task, however, the required
analysis seems to be much more complicated.

% conference papers do not normally have an appendix

% use section* for acknowledgement
\section*{Acknowledgment}
This work was supported by the German Research Foundation (DFG) under grant STA 864/7-1 and by the German Ministry of Research and Education (BMBF) under grant 01BU1224. The authors would like to thank J. Mohammadi, R.L.G. Cavalcante and M. Goldenbaum for helpful comments and discussions as well as the authors of \cite{MiMeSe11} for making available their implementation, which was used as a reference for the implementation of Alg. \ref{alg:alg1}.

\newpage
%\IEEEtriggeratref{2}
%\IEEEtriggercmd{\enlargethispage{-25in}}
\bibliographystyle{IEEEbib}
\bibliography{refs}

\begin{thebibliography}{10}

\bibitem{NaGa07}
B.~Nazer and M.~Gastpar,
\newblock ``Computation over multiple-access channels,''
\newblock {\em IEEE Transactions on Information Theory}, vol. 53, no. 10, pp.
  3498--3516, 2007.

\bibitem{GoSt13}
M.~Goldenbaum and S.~Stanczak,
\newblock ``Robust analog function computation via wireless multiple-access
  channels,''
\newblock {\em IEEE Transactions on Communications}, vol. 61, no. 9, 2013.

\bibitem{goldenbaum2010computing}
M.~Goldenbaum and S.~Stanczak,
\newblock ``Computing functions via {SIMO} multiple-access channels: {H}ow much
  channel knowledge is needed?,''
\newblock in {\em Acoustics Speech and Signal Processing (ICASSP), 2010 IEEE
  International Conference on}. IEEE, 2010, pp. 3394--3397.

\bibitem{boche2002iterative}
H.~Boche and S.~Stanczak,
\newblock ``Iterative algorithm for finding optimal resource allocations in
  symbol-asynchronous cdma channels with different sir requirements,''
\newblock in {\em Signals, Systems and Computers, 2002. Conference Record of
  the Thirty-Sixth Asilomar Conference on}. IEEE, 2002, vol.~2, pp. 1909--1913.

\bibitem{StHe03}
T.~Strohmer and R.~W. Heath,
\newblock ``Grassmannian frames with applications to coding and
  communication,''
\newblock {\em Applied and computational harmonic analysis}, vol. 14, no. 3,
  pp. 257--275, 2003.

\bibitem{LiHaCh07}
P.~Li, T.~J. Hastie, and K.~W. Church,
\newblock ``Nonlinear estimators and tail bounds for dimension reduction in
  $\ell_1$ using cauchy random projections,''
\newblock in {\em Learning Theory}, pp. 514--529. Springer, 2007.

\bibitem{Wi12}
A.~Winkelbauer,
\newblock ``Moments and absolute moments of the normal distribution,''
\newblock {\em arXiv preprint arXiv:1209.4340}, 2012.

\bibitem{WrNo99}
S.~J. Wright and J.~Nocedal,
\newblock {\em Numerical optimization},
\newblock Springer New York, 1999.

\bibitem{MiMeSe11}
B.~Mishra, G.~Meyer, and R.~Sepulchre,
\newblock ``Low-rank optimization for distance matrix completion,''
\newblock in {\em IEEE Conference on Decision and Control and European Control
  Conference (CDC-ECC)}, 2011.

\bibitem{HjGe07}
A.~Hjorungnes and D.~Gesbert,
\newblock ``Complex-valued matrix differentiation: Techniques and key
  results,''
\newblock {\em IEEE Transactions on Signal Processing}, vol. 55, no. 6, pp.
  2740--2746, 2007.

\bibitem{VaPi93}
C.~F. Van~Loan and N.~Pitsianis,
\newblock {\em Approximation with Kronecker products},
\newblock Springer, 1993.

\bibitem{Re09}
D.~Redmond,
\newblock {\em Existence and Construction of Real-Valued Equiangular Tight
  Frames},
\newblock Ph.D. thesis, University of Missouri-Columbia, 2009.

\end{thebibliography}

%\begin{appendices}
%\section{Derivation of $a^2$}
%
%
%\end{appendices}

\end{document}